\documentclass{aastex}
\usepackage{spr-astr-addons}
\usepackage{natbib}

\shorttitle {Mass outflow in presence of Compton cooling}
\shortauthors {Santanu Mondal, Sandip K. Chakrabarti $\&$ Dipak Debnath}
\begin{document}
\title  {Spectral signatures of dissipative standing shocks and mass outflow in presence of Comptonization around a black hole}
\author{Santanu Mondal\altaffilmark{1}}
\and
\author{Sandip K. Chakrabarti\altaffilmark{2,1}}
\and
\author{Dipak Debnath\altaffilmark{1}}
\email{santanu@csp.res.in, chakraba@bose.res.in, dipak@csp.res.in}
\altaffiltext{1}{Indian Centre For Space Physics, 43 Chalantika, Garia Station Road, Kolkata 700084, India}
\altaffiltext{2}{S.N. Bose National Center for Basic Sciences, JD-Block, Salt Lake, Kolkata,700098, India}


\begin{abstract}
Accretion flows having positive specific energy are known to produce outflows and winds which 
escape to a large distance. According to Two Component Advective Flow (TCAF) model, 
centrifugal pressure dominated region of the flow just outside the black hole horizon, with or without shocks,
acts as the base of this outflow. Electrons from this region are depleted due to the wind and consequently, energy 
transfer rate due to inverse Comptonization of low energy photons are affected. Specifically, it 
becomes easier to cool this region and emerging spectrum is softened. Our main goal is to show spectral 
softening due to mass outflow in presence of Compton cooling. To achieve this, we modify Rankine-Hugoniot 
relationships at the shock front when post-shock region suffers mass loss due to winds and energy loss due to inverse 
Comptonization. We solve two-temperature equations governing an accretion flow around a black hole which 
include Coulomb exchange between protons and electrons and other major radiative processes such as 
bremsstrahlung and thermal Comptonization. We then compute emitted spectrum from this post-shock flow. We 
also show how location of standing shock which forms outer boundary of centrifugal barrier changes with 
cooling. With an increase in disc accretion rate $(\dot{m_d})$, cooling is enhanced and we find that 
the shock moves in towards the black hole. With cooling, thermal pressure is reduced, 
and as a result, outflow rate is decreased. We thus directly correlate outflow rate with spectral state of the disc.
\end{abstract}



\section{Introduction}

The problem of origin and acceleration of jets and outflows in systems containing black holes has not yet been fully 
understood. The most super-massive black hole candidate known to date, namely, M87 has a very powerful jet 
and it appears to have a base of a few Schwarzschild 
radius \citep[][hereafter JBL99]{junor99}    
According to two component advective flow (TCAF) model of 
\citet[][hereafter CT95]{ct95} 
a black hole accretion consists of two major components: one is the usual Keplerian disc which 
becomes transonic only close to the black hole (inside the inner stable circular orbit) and other 
is a low angular momentum, sub-Keplerian flow which is presumably generated out of winds from the 
companion. This latter component is puffed up close to the black hole (a few tens of Schwarzschild radii)
and behaves like a Compton cloud. Relative importance of the rates of these two components determines 
the spectral state of emitted radiation. Furthermore, the same puffed up region is suggested 
(\citealt{c98}, hereafter C98; \citealt{c99}, hereafter C99) 
to be the source of matter for the outflows. Thus the JBL99 observation is in line with this suggestion 
of origin of jets. In other words, outflows do not emerge from entire accretion disc and they are formed from 
inner few tens of Schwarzschild radii. In case of stellar mass black hole candidates (e.g., GRO~J1655-40, 
GRS~1915+105, GX~339-4 etc.) jets and outflows are observed \citep[][and references therein]{mirabel99,mcClintock06}.
However, a similar measurement of size of base of jets is lacking due to low resolution 
of present instruments, though it is quite clear from observations that steady outflows are produced 
mainly when a black hole is in a hard spectral state 
\citep[][hereafter FBG04]{fender04}, 
namely, when the Compton cloud is present.
In passing, we may mention that the fact that a standard Keplerian disc, i.e., a \citet{ss73} 
disc supplies soft photons and produce black body spectrum in all models of black hole accretion 
is well established. Only difference from one model to another model is the nature of the Compton 
cloud which is known to inverse Comptonize these soft photons 
\citep[e.g.][]{haardt93,zdziarski03}. 
In TCAF scenario of CT95, a lower angular momentum flow which also 
produces winds and outflows from the inner, puffed up region, acts as the Compton cloud.

Many works are present in the literature on the origin of jets. \citet{cam86} 
and his group strongly suggest that magnetic field plays a major role in the production of 
the radio jets. Several authors \citep{bland74,tche11}
showed formation and collimation of jets through hydrodynamic and hydromagnetic processes 
where most of the power may directly come from the spin energy of the black hole or the disc.
On the other hand, recently it has been found that there is no evidence that jets are powered 
by black hole spin \citep{fender10}, 
at least in case of X-ray binaries and these authors speculate that this conclusion remains 
valid even for active galaxies and quasars. Second, if one assumes that the acceleration is 
from radiative processes \citep{chatto05} 
or some effects farther out, a supply of matter from the centrifugal barrier,  
from a few percent to a few tens of percent of the accretion flows 
(\citealt{mlc94}, hereafter MLC94; C88; C99; \citealt{cdc04}, hereafter CDC04; \citealt{ggc12}, hereafter GGC12),
is enough. Collimation may also be done by hoop stress 
\citep{konigl89, cb92} 
of toroidal fields escaping from accretion flow due to buoyancy effects 
\citep{cd94}. 
So it may not be essential to have strong and large scale aligned magnetic field in the disc 
itself. In the rotating wind solution of \citet{c86}, 
the flow was assumed to be coming out of the thick accretion disc 
(\citealt{pac80}, hereafter PW80; \citealt{c85}). 
However, MLC94 showed that post-shock region itself has all the properties of a thick 
accretion disc. It was shown also that, shocks are essential for driving outflows 
\citep{becker08}. 
C98 and C99 proposed that perhaps entire outflow is produced by this post-shock region, 
either steadily, as in a hard state, or episodically, when this region (commonly known as 
the CENtrifugal pressure dominated BOundary Layer or CENBOL) collapses due to enhance 
tension of toroidal magnetic field \citep{nandi01}. 
The outflow rate was found to be a function of compression ratio (ratio of post-shock  
and pre-shock densities) of the shock. This work was subsequently extended by adding angular 
momentum \citep[][hereafter D01]{dcnc01} 
to this flow and a similar conclusion was obtained. In these semi-analytical works, it became 
clear that when the shock is very weak (R $\sim$ 1) or very strong (R $\sim$ 4-7), the ratio 
of the outflow rate to the inflow rate $R_{\dot m}$ is low, but when the shock strength is
intermediate (say $\sim$ 2-3) $R_{\dot m}$ is very high and outflow rate could be as high as 
$20$\% of inflow rate. 

So far, however, no radiative process was explicitly included in solving governing hydrodynamic equations. Thus, the
effects of changing accretion rates of Keplerian component $(\dot{m_d})$ and sub-Keplerian 
component $(\dot{m_h})$ on the outflow rate are not studied from theoretical perspective. 
There are some studies which treat mass loss self-consistently, and yet, energy loss as 
parameters. \citet[][hereafter SC11]{sc10} 
used self-consistent mass loss but parametric cooling, and 
\citet[][hereafter MC13]{mc13} 
used self-consistent cooling while ignoring mass-loss altogether. They showed that parameter
space in which shocks may form, decreases due to cooling and mass loss effects. MC13 also 
showed that the presence of cooling due to Comptonization softens the spectra of emitted photons. 
These results indicate that there should be a direct correlation between spectral states and 
outflow rates. Indeed, it has been pointed out by \citet[][hereafter CC02]{cc02} 
that when matter leaves CENBOL, there is some evidence of softening and when cooler matter returns 
to CENBOL, there is evidence of hardening of spectra. So, it is all the more important to 
obtain solution of transonic flows which include Compton cooling and mass loss. Using this, 
we expect to answer the following questions: 
(a) is the cooling due to Compton scattering dynamically important? If so, does the 
shock location change significantly in its presence? 
(b) How would the result quantitatively be affected when mass loss from CENBOL is taken into account? 
(c) Would the CENBOL survive when disc rate ($\dot{m_d}$) is increased?
(d) How would mass loss rate depend on accretion rate? In outbursting sources, CENBOL is known to shrink 
on successive days as spectral state changes progressively from hard, hard-intermediate, soft-intermediate 
state to soft states \citep{van04,homan05,remillard06,dutta08,nandi12,dcnm08,dcn10,dcn13}.
The frequency of QPOs are also progressively increased in hard and hard-intermediate spectral states, 
become sporadically on and off in soft-intermediate spectral state and vanishes in soft state 
\citep[][hereafter NDC12, and references therein]{nandi12}. 
We believe that our result will throw some light on these phenomena.

In the next Section, we present governing equations and discuss procedures to solve these equations.  
We solve accretion flow equations using energy, baryon number and pressure balance conditions at 
shock front. For this, we write shock invariant quantities in terms of flow parameters.
In \S 3, we present flow variables as a function of disc accretion rate ($\dot{m_d}$), 
when halo accretion rate ($\dot{m_h}$) is fixed. We also present, how the allowed parameter space shrinks 
with increasing disc rate. Finally, in \S 4, we present summary of our work and concluding remarks 
regarding the cooling of the post-shock region, softening of spectrum and outflow rate.

\section {Governing equations and shock conditions}

\citealt{c89}, hereafter C89, presented a general transonic flow model called hybrid model. 
Modeling of a three dimensional flow around a back hole is done by assuming the flow to be 
rotating around the vertical axis. We assume the flow to be thin, axisymmetric, 
and non-dissipative in nature. Moreover, as in C89, the flow is transonic and thus advection dominated. 
To avoid integrating in a direction transverse to flow motion, we consider that the flow is 
in hydrostatic equilibrium in that direction. Governing equations are written down in equatorial plane of the disc where 
vertically averaged pressure and density at equatorial plane can be used.
This flow solution strictly depends on initial flow parameters, 
such as specific energy, specific angular momentum and accretion rate.
We begin with the assumption that non-dissipative, 
adiabatic, sub-Keplerian, inviscid matter is accreted on to a Schwarzschild black hole. The general 
relativistic effect is taken care of by Paczy'nski-Wiita (PW80) potential. Basic equations
are written on the equatorial plane of the accretion disc while radial momentum
balance equation which must be satisfied at the shock is vertically integrated \citep{mat84}.

Relevant hydrodynamic equations are already given in 
(C89; \citealt{c90a}, hereafter C90a; \citealt{c90b}, hereafter C90b; \citealt{c90c}, hereafter C90c) 
and we mention them here for the sake of completeness.

Radial momentum equation is given by:
$$
v\frac{dv}{dr} +\frac{1}{\rho} \frac{dP}{dr} - \frac{\lambda^{2}}{r^{3}} + \phi ' (r) = 0.
\eqno{(1)}
$$
where, $r$ is the radial distance and $-\phi ' (r)$ is the gravitational force due to the black hole.
Integrating this, we obtain specific energy of the accretion flow to be,
$$
\varepsilon = \frac{v^2}{2} + na^{2} + \frac{\lambda^{2}}{2r^{2}} - \frac{(r-1)^{-1}}{2},
\eqno{(2)}
$$
where, $P$, $\rho$ and $a$ are thermal pressure, density and sound speed, respectively, $\lambda$ 
is specific angular momentum,
$v$ is infall velocity, $n=\frac{1}{(\gamma-1)}$ is the polytropic index, $\gamma$ is the 
adiabatic index of the flow, $a = \surd(\frac{\gamma P}{\rho})$ and $\phi (r) = -\frac{(r-1)^{-1}}{2}$ (PW80).
The mass conservation equation is given by, 
$$
\dot{M} =  \rho  v rh(r).
\eqno{(3)}
$$
Here, $h(r)$ represents half-thickness of the flow at a radial distance $r$.
It is useful to rewrite this equation in terms of $v$ and $a$ in the following way:
$$
\dot{\cal{M}}= v a^q f(r), 
\eqno{(4)}
$$
where, $q=(\gamma+1)/(\gamma-1)$, $f(r)=r^{3/2}(r-1)$ for a vertical flow. Vertical flow implies that 
the disc is in hydrostatic equilibrium in vertical direction and therefore, local disc height is
obtained by equating pressure gradient force in vertical direction with the
component of gravitational force in that direction (C89). $\dot{\cal{M}}$ is the entropy 
accretion rate (first introduced in C89) which is conserved for an ideal flow but can vary in presence of shocks, where entropy is 
generated. Flow equations are made dimensionless considering units of
length, time and mass, i.e., $2GM_{bh}/c^2$, $2GM_{bh}/c^3$ and $M_{bh}$
respectively, where, $G$, $M_{bh}$ and $c$ represent universal gravitational
constant, mass of black hole and velocity of light respectively.

For radiative transfer, we remind the readers that soft photons produced by a Keplerian disc obey 
a multicolor black body spectrum coming from a standard (SS73) disc. We 
assume the disc to be optically thick due to free-free absorption, which is more important than opacity 
due to scattering. In this case the local emission is black body type with local surface temperature can be obtained
from SS73 prescription. In our case, Keplerian disc emits a flux $F_{ss}$ \citep{shappiro83} 
from the shock till the outer edge:
$$
F_{ss}=6.15\times10^{8} r^{-3}{\Im}\left (\frac{M_{bh}}{M_{\odot}}\right)^{-2}\dot{M} ~~ergs~cm^{-2}s^{-1}.
\eqno{(5)}
$$
Here, $\Im = 1-(3/r)^{1/2}$. In above equations, mass of the black hole $M_{bh}$ is measured in units of mass of 
Sun ($M_\odot$), and disc accretion rate $\dot{M_d}$ is in units of $~gm~sec^{-1}$. 
For concreteness, we choose a stellar mass black hole: $M_{bh} = 5 M_{\odot}$, ${\dot m}_{d}$=$\frac{{\dot M}_{d}}{\dot{M}_{Edd}}$   
and ${\dot m}_{h}$=$\frac{{\dot M}_{h}}{\dot{M}_{Edd}}$, where, ${\dot M}_{h}$ and $\dot{M}$$_{Edd}$ are 
halo accretion rate and Eddington rate for rest of the paper. We use these quantities 
in modified Rankine-Hugoniot conditions to obtain shock invariant quantity 
(C89) to be satisfied between pre- and post-shock region in presence of cooling and mass loss (outflows).

\subsection{Shock condition and shock constant in presence of outflow (mass loss)}

Angular momentum of the halo causes centrifugal barrier to form a shock (C89, C90a,b). Post-shock 
region (CENBOL), which acts as a boundary layer of the black hole, is hot since kinetic energy of pre-shock 
flow is essentially converted into thermal energy. As a result, it is puffed up and it intercepts soft photons 
from the Keplerian disc (CT95) and reprocesses them through Compton scattering. If intercepted 
soft-photon number is high enough, CENBOL is cooled down due to inverse Comptonization (CT95, MC13) 
and the spectrum become softer. Due to loss of energy at the shock front, shock condition for the 
black hole accretion flow is, 
$$
\varepsilon_{+}=\varepsilon_{-} - \Delta \varepsilon,
\eqno{(6a)}
$$
where, $\Delta \varepsilon$ is energy loss due to Comptonization. This is basically a function of number 
density of electrons and number of low energy photons and can be calculated for different accretion rates 
of the flow. Subscripts `-' and `+' denote pre-shock and post-shock quantities, respectively.
Baryon number conservation of the flow gives,
$$
\dot{M_{+}}=\dot{M_{-}}(1-R_{\dot m}),
\eqno{(6b)}
$$
Here, $R_{\dot m}$ denotes the ratio of outflow and inflow rates. Because the gas puffs up, Rankine-Hugoniot 
conditions \citep{landau59} 
have to be modified so that only vertically integrated pressure and 
density are important. This modification was first carried out in C89, where pressure balance condition was
written as:
$$
W_{+} + \Sigma_{+}v_{+}^2 = W_{-} + \Sigma_{-}v_{-}^2.
\eqno{(6c)}
$$
Here, $W$ and $\Sigma$ are pressure and density, integrated in vertical direction \citep{mat84}.
To get location of a shock, it is better to express flow parameters in terms of some invariant quantities
which remain the same on either side of the shock (C89).
For this, we rewrite equations (6a), (4) and (6c) in terms of Mach number $M=\frac{v}{a}$ of the flow,  
$$
\frac{1}{2}M_{+}^2 a_{+}^2 +\frac{a_{+}^2}{\gamma-1} = \frac{1}{2}M_{-}^2 a_{-}^2 +\frac{a_{-}^2}{\gamma-1} - \Delta \varepsilon,
\eqno{(7a)}
$$
$$
\dot{\cal{M}_{+}}= M_{+} a_{+}^\nu f(x_{s}),
\eqno{(7b)}
$$
$$
\dot{\cal{M}_{-}}= M_{-} a_{-}^\nu f(x_{s}),
\eqno{(7c)}
$$
and
$$
\frac{a_{+}^\nu}{\dot{\cal{M}_{+}}}\left[\frac{2\gamma}{3\gamma -1}+\gamma M_{+}^2\right]=
\frac{a_{-}^\nu}{\dot{\cal{M}_{-}}}\left[\frac{2\gamma}{3\gamma -1}+\gamma M_{-}^2\right],
\eqno{(7d)}
$$
where, $\nu=\frac{3\gamma -1}{\gamma-1}$ and $x_{s}$ is the location of the shock. After some algebra, we obtain
a Mach number relation which connects pre-shock and post-shock quantities in the same way as was done for
non-dissipative flow without mass loss (C89). This relation is (SC10):
$$
           \frac{[M_{+}(3\gamma-1)+(\frac{2}{M_{+}})]^2[1-R_{\dot{m}}]^2 }{2+(\gamma-1)M_{+}^2}=
           \frac{[M_{-}(3\gamma-1)+(\frac{2}{M_{-}})]^2}{2+(\gamma-1)M_{-}^2-\zeta} ,
\eqno{(8a)}
$$
where, $\zeta=\frac{2 \Delta \varepsilon (\gamma-1)}{a_{-}^2}$.
$M_{-}^2$ can be evaluated using Eqs. 7(a-c):
$$
M_{-}^2=\frac{\left(\frac{\dot{\cal{M}_{+}} M_{-}}{\dot{\cal{M}_{-}} M_{+}}\right)^{\frac{1}{4}}\left[2+(\gamma-1)M_{+}^2\right]-2}{(\gamma-1)}
+\frac{2\Delta \varepsilon}{a_{-}^2}. 
\eqno{(8b)}
$$
However, if the flow were one dimensional (cylindrical with constant height, or, of conical shape having constant
wedge angle), then a different Mach number relation would have to be used since vertical integration would 
not be necessary (C90a; \citealt{dcm10}, hereafter D10; \citealt{sc10}, hereafter SC10). 
In this case, Mach Number relation becomes:
$$
                    \frac{\left[M_{+}\gamma+(\frac{1}{M_{+}})\right]^2\left[1-R_{\dot{m}}\right]^2}{2+(\gamma-1)M_{+}^2}=
                    \frac{[M_{-}\gamma+(\frac{1}{M_{-}})]^2}{2+(\gamma-1)M_{-}^2-\zeta}
\eqno{(9)}
$$
In our solution we use each side of Eq. 8a as an invariant quantity across the shock.

It is to be noted that our energy loss is gradual and not instantaneous. Ideally, this feature should have been incorporated in the 
shock condition itself. However, because cooling is due to Comptonization which is highly non-local, 
it is impossible to express the loss as a function of radial distance.  In order to obtain the 
net loss, we sum the emitted spectrum over 
the whole frequency range, which can be written as:
$$\sum_{i=\nu_{l}}^{\nu_{u}} \nu_i F_{Comp}(i),$$ where $\nu_{l}$ and $\nu_{u}$ are the lower 
and upper limits of the frequency. We convert this frequency unit to energy [$EF(E)$] unit. 
We multiply this by the surface area 
$A_{surf}=4 \pi (X_s-X_{in})^2$, of the CENBOL. This  gives us the cooling rate. Here, $X_{in}$ is the inner region of the disc. This energy is assumed to be released 
instantaneously immediately after the shock in order to obtain the Mach number relation.
 
\subsection {Ratio of outflow to inflow rate}

We assume that matter in pre-shock region is cool enough so that we may ignore thermal pressure.
From energy conservation, velocity of matter will be, 
$$
v(r)=\left[\frac{1}{r-1}-\frac{\lambda^{2}}{r^{2}}\right].
\eqno{(10)}
$$
At the shock location, we can calculate compression ratio using mass conservation equation which is given by,
$$
R=\frac{\Sigma_+}{\Sigma_-}=\frac{h_+(x_s) \rho_{+}(x_{s})} {h_-(x_s) \rho_{-}(x_{s})} = \frac {v_{-}}{v_{+}}.
\eqno{(10a)}
$$
To calculate outflow rates in terms of known quantities, we assume that in pre-shock region, thermal pressure 
is very small compared to ram pressure. Thus from Eqn. 6c, for vertical equilibrium model (D01),
$$
W_{+}(x_{s})=\frac {R-1}{R}\Sigma_{-}(x_{s})v^{2}_{-}(x_{s}),
\eqno{(10b)}
$$
where, $x_{s}$ is the location of the shock (in Schwarzschild radius $r_s =2GM_{bh}/c^{2}$). 
Isothermal sound speed $C_s$ in post-shock region is obtained from (D01),
$$
C^2_{s}=\frac {R-1}{R^2}v^{2}_{-}
=\frac {1}{f_{0}}\left[\frac {x^{2}_{s}-\lambda^2(x_{s}-1)}
{x^{2}_{s}(x_{s}-1)}\right],
\eqno{(10c)}
$$
where, $f_{0}=\frac {R^2}{R-1}$. At CENBOL, adiabatic sound speed is, $a_s^2 = \gamma C_s^2$.
From adiabatic equation of state of flow, ratio of density at 
critical point of outflow and density at the CENBOL is given by,
$$
\frac{\rho_c}{\rho_s} = \left[\frac{a_c^{2}}{a_s^{2}}\right]^{n}.
\eqno{(10d)}
$$
Outflow rate is given by:
$$
{\dot M}_{o}= \Theta_{o} \rho_{c}v_{c} x_c^{2},
\eqno{(10e)}
$$
where, $\Theta_{o}$ is solid angle subtended by outflow, subscript ``c" denotes quantities at  
critical point of outflow. Using 
Eq. 10(a-e) and after some algebra (C90a, D01) one obtains ratio of outflow rate to inflow rate as (SC10): 
$$
R_{\dot m}=\frac{\Theta_{o}}{\Theta_{i}}\frac{\left[\frac{2n+1}{2n}
(\frac{1}{5x_{c}}-\frac{2\lambda^{2}}{5x_{c}^2})\right]^{\eta}}
{\left[\frac{(n+1)}{n}C^2_{s}\right]^{n}}R\sqrt{\frac{n}{\eta}}[C^2_{s}f_0]^{-\frac{1}{2}}\left(\frac{x_{c}}{x_{s}}\right)^{2}
\eqno{(10f)}
$$
where, $\Theta_{i}$ is solid angle subtended by inflow and $\eta=\frac{(2n+1)}{2}$. 
For a relativistic flow $n=3$ and $\frac{\Theta_{o}}{\Theta_{i}}=0.1$ for reference.
Height of the shock which we have calculated self-consistently in this paper is 
$h_{s}=\left(\frac{2}{\gamma}\right)^{1/2}a_{s}x_{s}^{1/2}(x_{s}-1)$, where $a_{s}$ is sound 
speed at post-shock region (CDC04).
Now we compute outflow rate self-consistently in presence of Compton cooling.

\subsection{Radiative Processes }

Detailed radiative processes in accretion flows are discussed in \citet{tl95} 
and CT95. We briefly discuss them here for the shake of completeness. Accreted matter becomes hotter 
due to geometrical effects and electrons loss energy due to bremsstrahlung and Comptonization of the soft 
photons emitted by the Keplerian disc. Electrons gain energy from protons due to Coulomb interaction and
as a result, protons are cooled down. Energy equation which protons and electrons obey in post-shock 
region is given by,
$$
\frac{\partial(\varepsilon+\frac{P}{\rho})}{\partial r}+(\Gamma - \Lambda)=0,
\eqno{(11)}
$$
where, $\varepsilon$ is given by Eq. (2) and $\Gamma$ and $\Lambda$ are heating and cooling terms, respectively.
If electron temperature $T_e$ is high enough, $kT_e > m_e c^2$, we use $\gamma=4/3$, otherwise, we use $\gamma= 5/3$.
Density of post-shock region is $\rho=\frac{R \dot M_h +\dot M_d}{4\pi r^{2}v(r)}$, 
where $R$ is the compression ratio of the flow, $r$ and $v(r)$ 
are radial distance and velocity of the flow. Accretion rates are obtained from pre-shock values.  
Geometric factor $4\pi$ arises because we assume spherical shape of CENBOL. 
Thus density of post-shock region
is a mixture of both disc and halo components of the flow.
We ignore synchrotron cooling for the time being. Synchrotron cooling will introduce more seed photons and 
is expected to cool down CENBOL region even farther.
During Comptonization, energy exchange takes place through scattering of photons off free electrons. 
Average energy exchange per scattering is given by ($h\nu, kT_e \ll m_e c^2$), 
$$
         \frac{\Delta \nu}{\nu} = \frac{4kT_e - h\nu}{m_e c^2}.
\eqno{(12)}
$$
When $h\nu \ll kT_e$, photons gain energy due to the Doppler effect by scattering with electrons having significant bulk motion.
Power-law distribution of energy which we use in our calculation is given by,
$$
F_{\nu} \propto \nu^{-\alpha}.
\eqno{(13)}
$$

Though we get post-shock temperatures of proton and electrons using two temperature 
equation at in CT95. Then from these, an average temperature was obtained for spectral study.
we assume that the outflow, at least up to the                
sonic surface, has the same temperature as that of the average temperature of CENBOL. This is justifiable as 
density is higher in sub-sonic region and it is also illuminated by radiation from 
CENBOL. So cooling due to adiabatic expansion of outflow is assumed to be compensated by heating
from CENBOL. After flow becomes super-sonic this condition need not be justified.

\subsection{Solution procedure}
Detailed discussion on how to obtain self-consistent solution with Comptonization has been presented 
already in MC13. In this paper, we obtain solutions after inclusion of outflows. 
In a given run, we assume a Keplerian disc rate ($\dot{m}_d$) and a fixed sub-Keplerian halo rate 
($\dot{m}_h$) with a given set of specific energy ($\varepsilon$) and angular momentum ($\lambda$).
To include cooling and mass outflow in coupled radiative hydrodynamics code, we follow these steps: 
(i) As the first step of iteration, we calculate shock location, inner sonic point, and outer sonic point 
from the hydro-code following C89 without assuming any radiative transfer.
(ii) Assuming emission from Keplerian disc to be same as that from a standard Shakura-Sunyaev disc, 
cooling (heating) of electrons are computed using prescription of CT95. This gives us an average electron 
temperature and Comptonization enhancement factor.
(iii) Energy loss by electrons $(\Delta \varepsilon)$ is calculated by integrating emergent spectrum.
(iv) Albedo $A_{\nu}$ is calculated to determine fraction of CENBOL flux scattered from each radius of 
the disc. The rest $\cal B_{\nu}$=$(1-A_{\nu})$ is assumed to be absorbed by the disc and is re-radiated 
as a black body spectrum at a higher temperature. Since our model, by definition, includes the reflected component, we need not 
add any other reflected component as in often done in other models 
\citep[e.g.][]{garcia11}. 
(v) While calculating in dissipative regime, in the first step, value of $R_{\dot m}$ is assumed 
to be zero. When an energy loss is considered, we take new energy as initial condition and calculate 
new inner sonic point, which also gives a new shock location using modified Rankine-Hugoniot shock 
condition. These new parameters give $R_{\dot m}$, which is used as input of next iteration. In this 
way, we repeat steps (i) - (v) until the spectrum converges. 
At convergence, we get final shock location, final sonic point locations, outflow rate,  
emergent spectrum, spectral index and most importantly, a complete solution of transonic flow 
self-consistently, totally self-consistently. 
(vi) Whole procedure is then repeated for every flow parameter for every pair of disc and halo 
accretion rates to check which region of the parameter space allows shock formation.

\section{Results and Discussions}

Main concern in this paper is to study how emitted spectrum is affected when both mass loss and energy loss 
are present in an accretion flow and also to study region of parameter space which can produce shocks and outflows in presence of such a loss.
Close to a black hole, matter is compressed and heated up due to geometric compression and it leaves the disc
between centrifugal barrier and inner boundary. We obtained modified parameter space which allows shock 
formation in self-consistent transonic flow in presence of mass outflow and thermal Comptonization. In 
Figure 1, we show variation of allowed parameter space (spanned by specific energy and angular momentum) 
for different values of accretion rates for which outflows are present. Solid curve is for a non-dissipative 
flow. Dashed curve is drawn for accretion rate $\dot{m}_d=0.04$ and dash-dotted curve is drawn for accretion 
rate $\dot{m}_d=0.15$ in units of Eddington rate. Sub-Keplerian halo rate is $\dot{m}_h=1$. 
This shows that parameter space decreases self-consistently with increasing accretion rate. 
Parameter space decreases significantly from low angular momentum end because of drop of pressure 
due to mass loss and cooling effects. This implies that, shocks can form only if angular momentum is significant 
and close to marginally stable value. This also implies that as spectral state becomes softer, formation of 
outflows becomes impossible, a fact already confirmed from observations (FBG04) and numerical simulations (GGC12). 
\begin{figure}
\vspace {-0.3cm}
\centering{ 
\includegraphics[height=5.5truecm,angle=0]{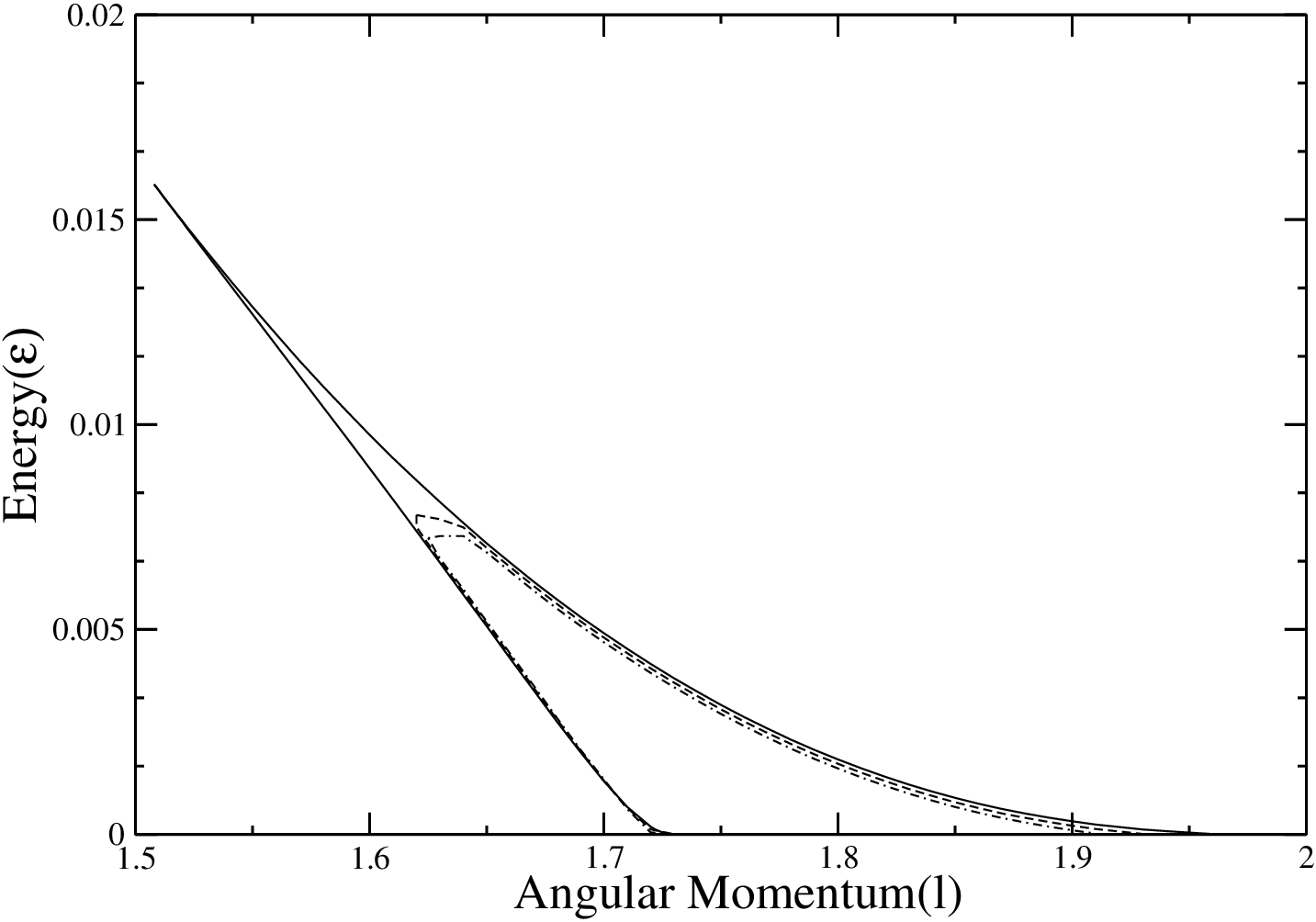}}
\caption{Allowed parameter space in presence of mass loss and thermal Comptonization. Solid, dashed and dash-dotted curves are 
for non-dissipative flows (i.e., $\dot{m}_d=0.0$), for $\dot{m}_d=0.04$ and for $\dot{m}_d=0.15$ respectively. With increase in accretion rate,
number of soft photons from a Keplerian disc is increased and thus amount of cooling ($\Delta \varepsilon$ ) increases. 
This shrinks down the parameter space responsible for production of mass outflows.}
\label{fig1}
\end{figure}

\begin{figure}
\vspace {0.4cm}
\vbox{
\centering{
\includegraphics[height=5.5truecm,angle=0]{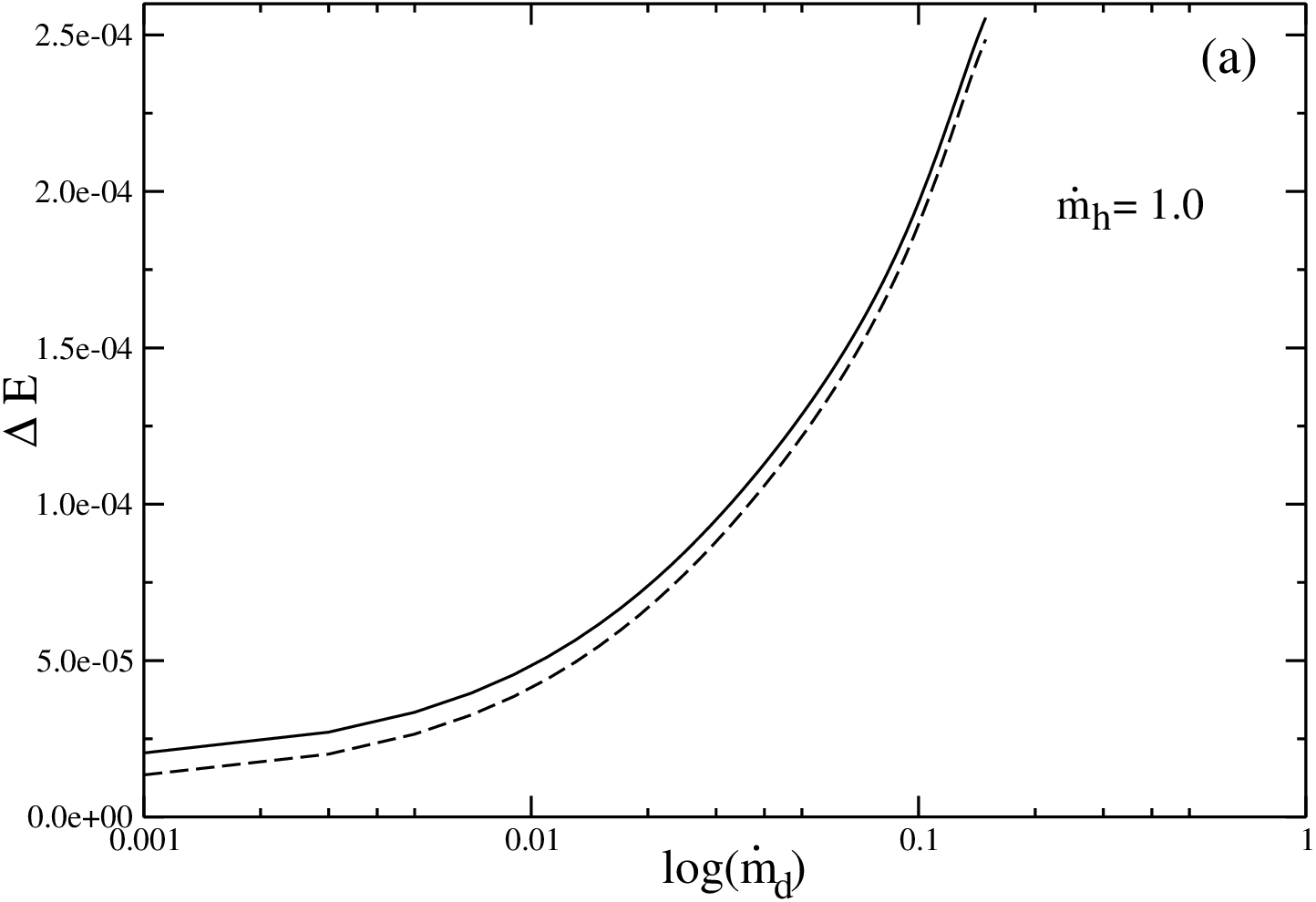} \vskip 1.0cm
\includegraphics[height=5.5truecm,angle=0]{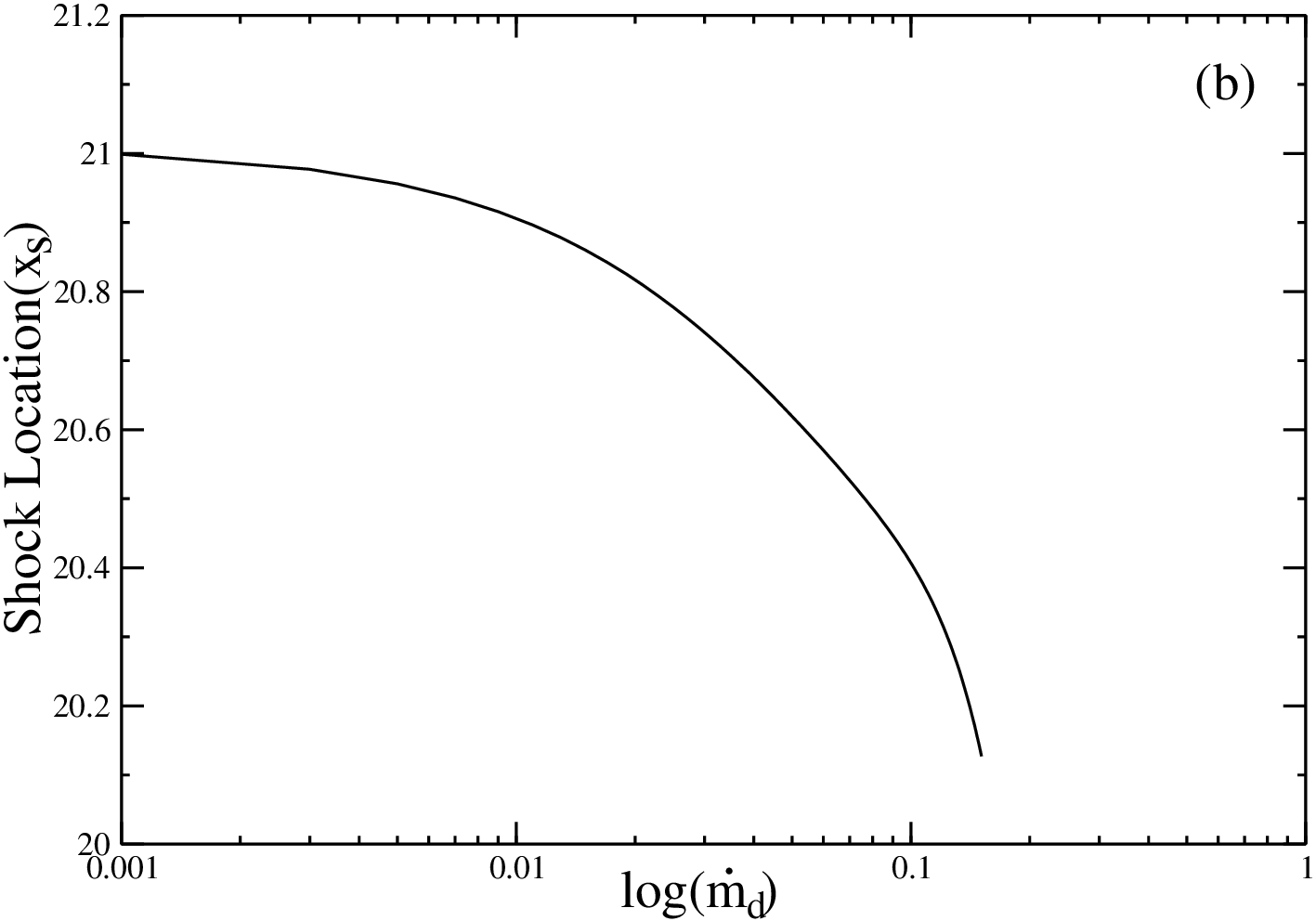}}
\caption{Variation of (a) lost energy $\Delta \varepsilon$ due to cooling from post-shock region due to Comptonization 
and (b) shock location as a function of disc accretion rate ($\dot{m}_d$). Clearly, $\Delta \varepsilon$ increases with 
$\dot{m}_d$, when mass loss ($R_{\dot{m}}$) is present.}}
\label{fig2ab}
\end{figure}
Figures 2(a-b) show that if we increase accretion rate ($\dot{m}_d$), (a) cooling increases and (b) shock 
moves toward the black hole. In both cases initial flow has identical set of specific energy and specific 
angular momentum ($\varepsilon=0.0021$, $\lambda=1.74$) and mass loss rate from CENBOL for 
halo rate $\dot{m}_h = 1$ and mass of black hole $M_{bh}=5M_\odot$. 
Thermal Comptonization becomes more effective and cooling of 
electrons is more efficient when Keplerian rate rises. As a result, pressure in post-shock region 
drops and also rate of outflow decreases as post-shock pressure 
is not sufficient to drive matter out from CENBOL. 
Shock moves toward the black hole to satisfy modified Rankine-Hugoniot condition.
\begin{figure}
\vspace {-0.1cm}
\vbox{
\centering{
\includegraphics[height=5.5truecm,angle=0]{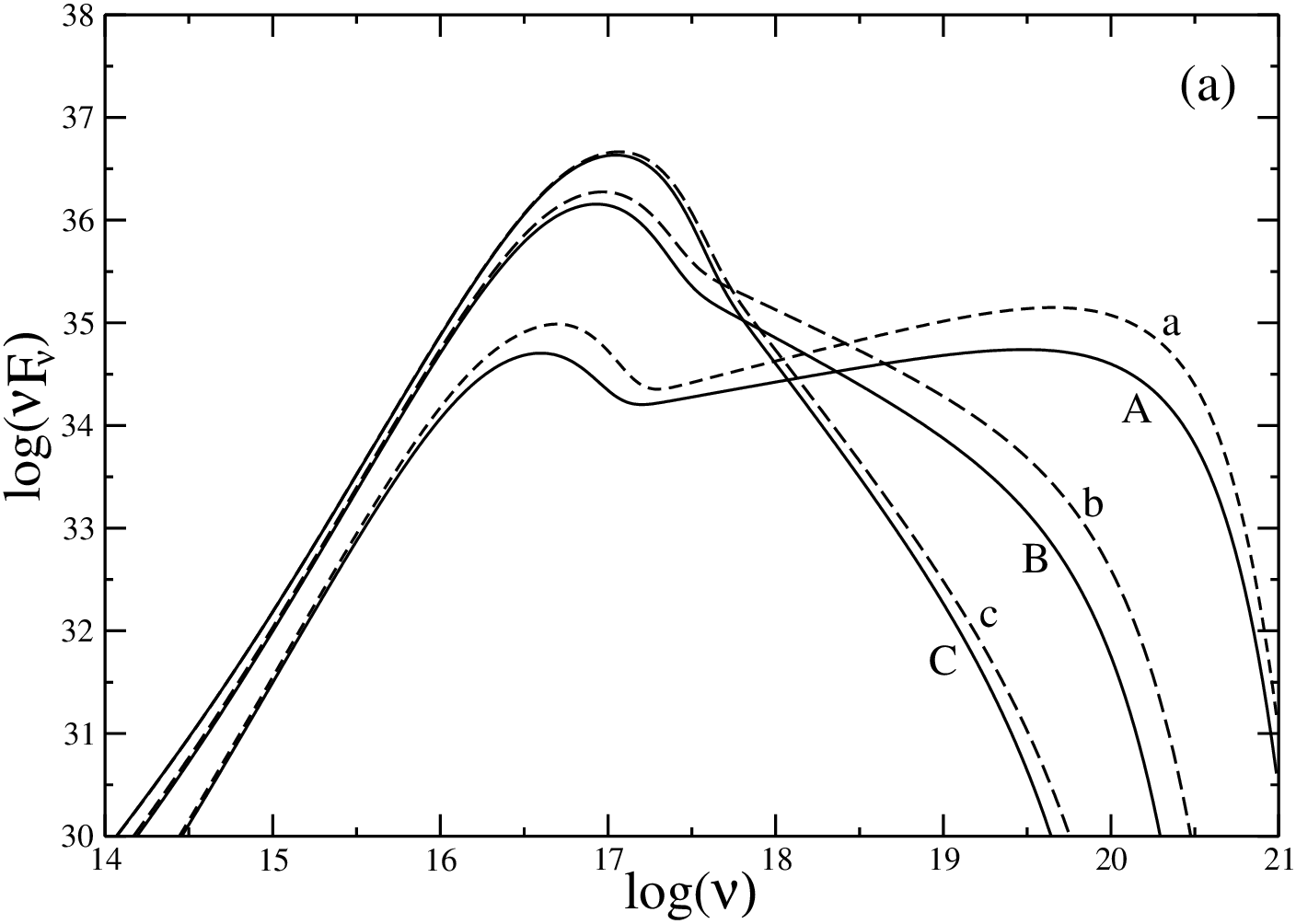}\vskip 1.0cm
\includegraphics[height=5.5truecm,angle=0]{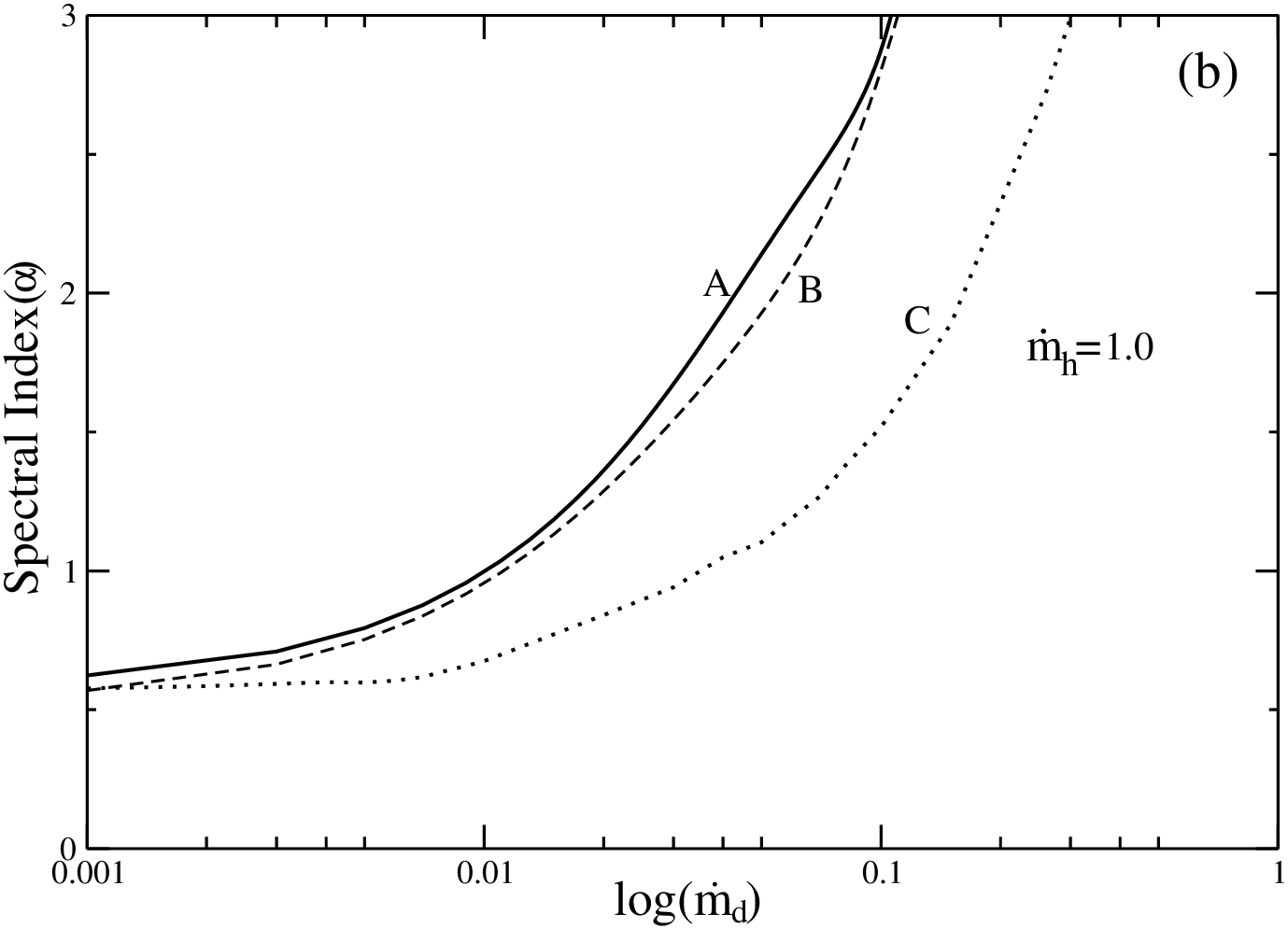}}
\caption{(a) Variation of spectrum for different accretion rates, keeping 
halo rate ($\dot{m}_h = 1$) fixed for the set of initial (non-dissipative) energy ($\varepsilon) = 0.0021$ 
and angular momentum ($\lambda) = 1.74$ of the flow. 
Solid curves show softening of spectra for $\dot{m}_d = 0.001$, $0.04$, and $0.15$ respectively when 
mass loss and Compton cooling are present. Dashed curves show spectra when only cooling is present.
(b) Variation of spectral index ($\alpha$) with accretion rate ($\dot{m}_d$) for halo rate $1$, when only cooling 
(dashed curve, MC13) and both mass loss and cooling (solid curve) are present. Dotted curve shows variation 
when non-dissipative flow (CT95) is assumed.} }
\label{fig3ab}
\end{figure}

In Fig. 3a, we show variation of energy spectrum with increase of Keplerian disc accretion rate, when halo rate 
$\dot{m}_h = 1$, initial energy $\varepsilon = 0.0021$ and angular momentum of the flow $\lambda = 1.74$ are 
fixed. Here, both Compton cooling and mass loss are considered self-consistently in post-shock region. Solid 
curves show spectra for $\dot{m}_d =$ (A) $0.001$, (B) $0.04$ and (C) $0.15$ respectively when both types of 
dissipation occur in CENBOL. Clearly, spectrum becomes softer for higher values of $\dot{m}_d $ as it 
increases number of injected soft photons which cool down CENBOL faster and also decreases radiation pressure 
at base of the jet. As a result, outflow rate also decreases. Dashed curves (a, b, c) show variation of energy 
spectrum when only Compton cooling is present. From this result, we conclude that spectrum softens quickly when 
both mass loss and cooling are present. Cooler CENBOL becomes smaller in size (reduction of sound speed reduces 
the vertical height of CENBOL). This is reflected in the value of spectral index. In Fig. 3b, we show variation 
of spectral index ($\alpha$) with accretion rates ($\dot{m}_d$). Since spectra are softer, curves are shifted 
upward in general. For example, for $\dot{m}_h = 1.0$, shifting is shown from C-A for different values of 
accretion rates. Dotted curve is for non-dissipative flow, dashed curve for dissipation by Compton cooling only 
and solid curve is for dissipation when mass loss and cooling are present. At a very low disc rate, 
spectral index is very insensitive to disc rate. However, since thermal Comptonization becomes effective
when CENBOL optical depth is above $\sim 2/3$, all of a sudden, cooling and mass loss take effect 
and soft state is formed. This phenomenon was conjectured to be responsible for burst-on and burst-off states 
of GRS 1915+105 \citep[][hereafter CM00]{cm00}. 

\begin{figure}
\vspace {0.0cm}
\centering{
\includegraphics[height=5.5truecm,angle=0]{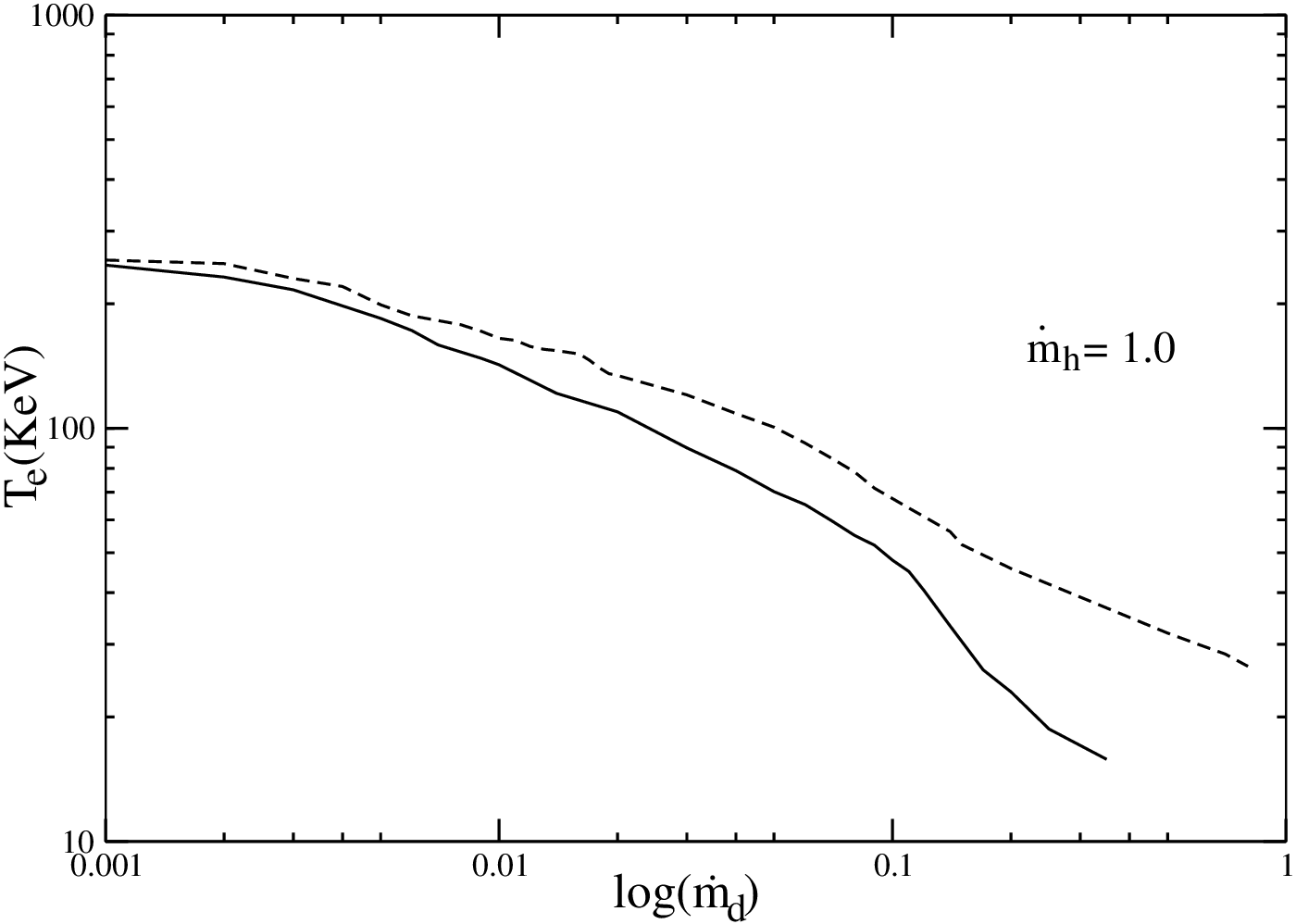}}
\caption{Variation of mean electron temperature (in KeV) of CENBOL in presence of Comptonization as a function 
of disc accretion rate (X-axis) for a set of energy and angular momentum when halo rate $\dot{m}_h=1$ with 
(solid curve) and without (dashed curve (MC13)) mass loss taken into account.}
\label{fig4}
\end{figure}

In Fig. 4, we show variation of mean electron temperature (in KeV) in presence of thermal Comptonization 
as a function of disc accretion rate (X-axis) for an initial set of energy ($0.0021$) and angular momentum 
($1.74$) when halo rate ($\dot{m}_h$) is $1$. Solid curve shows temperature variation of CENBOL with disc 
accretion rate when mass ejection and Compton cooling are present. As the hot CENBOL produces outflowing 
jet (C99; CC02) 
a large amount of thermal energy escapes from CENBOL. CENBOL becomes cool faster compared to the case 
when only Compton cooling takes place. Dashed curve shows same variation for flow with same parameters, 
when only Compton cooling is present (MC13). As in spectral index, temperature remains constant 
with disc accretion rate as long as CENBOL optical depth is not high enough. For higher accretion rates, 
CENBOL cools catastrophically (as also shown in CT95, MC13) and outflow rate decreases. 

\begin{figure}
\vspace{0.0cm}
\centering{
\includegraphics[height=5.5truecm,angle=0]{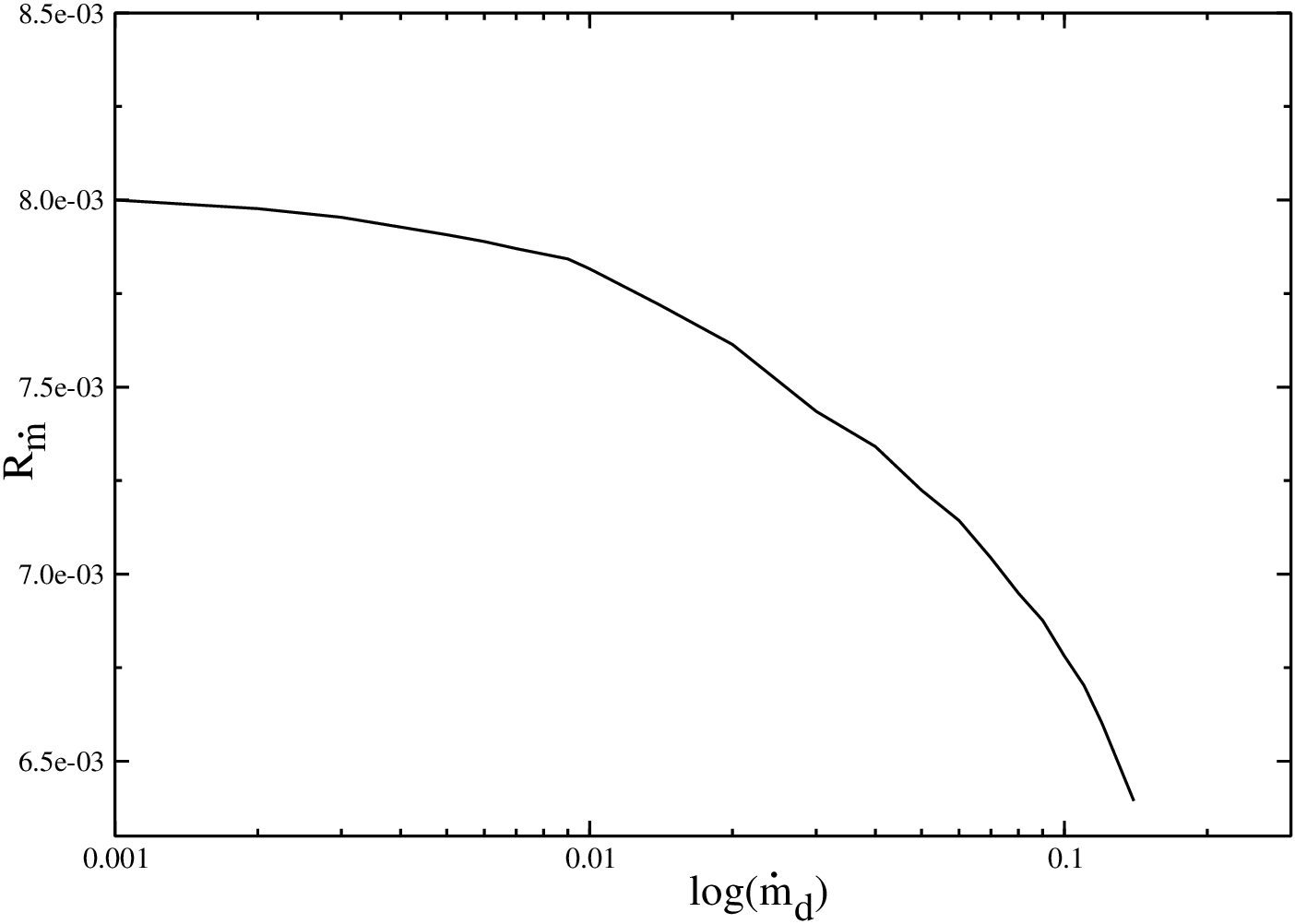}}
\caption{Variation of ratio of outflow and inflow rates $R_{\dot{m}}$ as a function of disc accretion rate.
As accretion rate increases, ratio of outflow and inflow rates decrease. See text for details.}
\label{fig5ab}
\end{figure}
In Fig. 5, we show, as an illustration, that $R_{\dot{m}}$ decreases as accretion rate increases. If we 
choose different flow parameters, in particular with higher initial flow energy and angular momentum, 
rate would be much higher, typically tens of percent of inflow. In our solution, we also see that for 
energy $0.00316$ and angular momentum $1.75$, value of outflow rate is $\sim$7.0\% .
\begin{figure}
\vspace{0.65cm}
\centering{
\includegraphics[height=5.5truecm,angle=0]{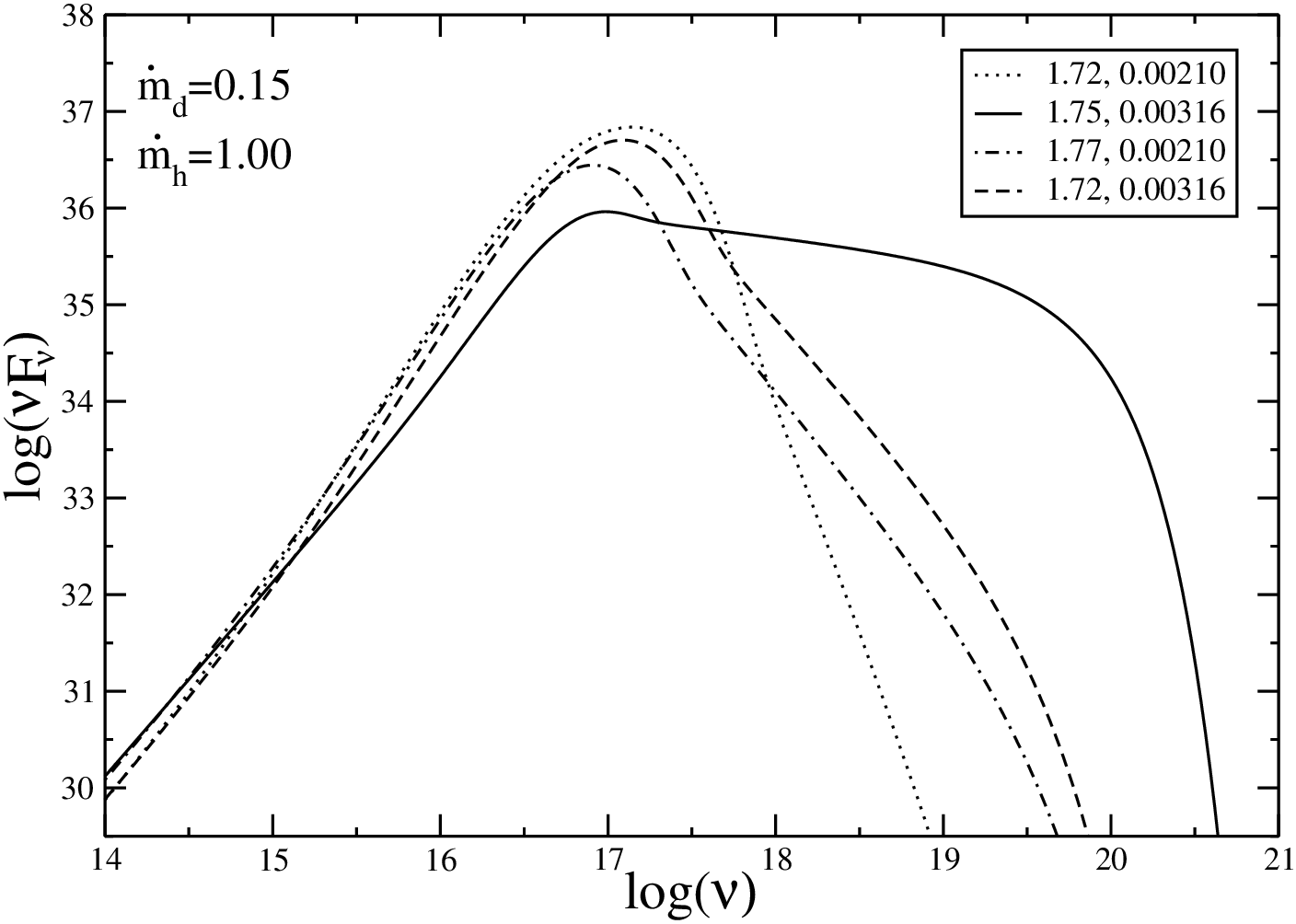}}
\caption{Variation of spectral shape for different sets of energy and angular momentum of the flow. See text for details.}
\label{fig6}
\end{figure}
Fig. 6 shows spectral variation for a different set of energy ($\varepsilon$) and 
angular momentum ($\lambda$) of flow. Dotted curve shows the variation when angular momentum is 
low ($1.72$). For this set of flow parameters, system is in soft state (size of the CENBOL is very small) 
and outflow rate is also low, $\sim0.63$\%. It is clear from our analysis
that spectral index and outflow rate change significantly with change in energy 
and angular momentum of the flow, when disc rate ($0.15$) and halo rate ($1.0$) are kept fixed.
Outflow is significant in hard states (C95; C99; \citealt{dcnc01}) when Keplerian rate is lower.
However, increase in accretion rate brings the system into a soft state and radiation pressure at 
base of jet is not sufficient to drive out matter from CENBOL. As a result, outflow rate decreases.  

\section{Summary and Conclusions}

In this paper, we study spectral properties of a transonic flow around a black hole when both mass outflow 
and energy loss due to thermal Comptonization are present. We also study formation of standing shocks  
in this generalized scenario, which was not attempted so far. Earlier (C99, D10, MC13) 
shock was found to be present even when post-shock region is losing energy. However, lose was modeled 
parametrically. In contrast, here, we study formation of outflows when post-shock region is cooling down due to Comptonization. Hot electrons 
in post-shock region (CENBOL acting as the Compton cloud) of a low angular momentum and low viscosity flow (halo) 
which surrounds Keplerian disc interact with these seed photons and energize them through inverse 
Comptonization to produce hard X-rays. In this situation as Keplerian disc accretion rate is increased, 
soft photon flux goes up, and Compton cloud is cooled down resulting in softening the spectrum (CT95, MC13). 
There is not enough thermal pressure in CENBOL to generate significant outflows. We demonstrated that an increase 
in disc accretion rate increases cooling in post-shock region resulting in reduction in outflow rate. 
Our results indicate that spectral index variation changes significantly when mass outflow is present 
along with thermal Comptonization in the CENBOL. We also show that at higher Keplerian rates, nature of 
variation in spectral index remains almost the same even in presence of cooling [curves (A) and (B) in 
Fig. 3b]. With outflows included, effects are more dramatic. In Fig. 3a, we see that effective change in 
spectra (C) is lesser as compared to changes in (A) and (B). We study parameter space in which standing 
shocks in a Comptonized transonic flow (including outflows) form. We find that area of parameter space 
shrinks with increase of cooling effects.

\citet{fender10}, 
discussed a relationship between spectral state and mass outflows and specifically showed that there 
was no evidence that spin of black hole plays a major role in producing outflows. They suspected 
some other factors. Our self-consistent solution strongly suggests that this unknown factor is possibly 
nothing but the properties of post-shock region, or CENBOL, which is the prime driver of all outflows and its properties are
decided mainly by angular momentum. CENBOL not only produces observed hard X-rays (CT95, MC13), it also supplies matter for jets and outflows 
(C99, SC10). Oscillation of CENBOL is known to produce QPOs observed in black hole candidates 
(\citealt{msc96}, CM00). 

Unlike earlier works (C99, D01, D10, SC11), present paper produces self-consistent transonic 
solution which couples both radiative transfer and hydrodynamic processes. Here cooling is neither 
a parameter nor an assumption. It is estimated directly from emitted spectrum. Shock height,
shock temperature and outflow rates are also obtained from our hydrodynamic solution. Inner boundary of Keplerian 
disc is neither a parameter nor a constant. It changes according to shock condition. 
Our present study puts this model in a firm footing, since we
prove that earlier results also remain valid in presence of   mass and energy losses studied parametrically.

\section{Acknowledgment}

Santanu Mondal acknowledges CSIR fellowship for this work. 


\end{document}